\pdfoutput=1
\RequirePackage{ifpdf}
\ifpdf % We are running pdfTeX in pdf mode
\documentclass[pdftex]{sigma}
\else
\documentclass{sigma}
\fi

\numberwithin{equation}{section}

\begin{document}

\allowdisplaybreaks

\newcommand{\arXivNumber}{1602.07927}

\renewcommand{\PaperNumber}{047}

\FirstPageHeading

\ShortArticleName{Nonstandard Deformed Oscillators from Deformed Heisenberg Algebras}

\ArticleName{Nonstandard Deformed Oscillators\\ from $\boldsymbol{q}$- and
$\boldsymbol{p,q}$-Deformations of Heisenberg Algebra}

\Author{Alexandre M.~GAVRILIK~$^\dag$ and Ivan I.~KACHURIK~$^{\dag\ddag}$}

\AuthorNameForHeading{A.M.~Gavrilik and I.I.~Kachurik}

\Address{$^\dag$~Bogolyubov Institute for Theoretical Physics, 14-b
Metrolohichna Str., Kyiv, 03680 Ukraine}
\EmailD{\href{mailto:omgavr@bitp.kiev.ua}{omgavr@bitp.kiev.ua}, \href{mailto:kachurik@bitp.kiev.ua}{kachurik@bitp.kiev.ua}}

\Address{$^\ddag$~Khmelnytskyi National University, 11 Instytutska Str., Khmelnytskyi, 29016 Ukraine}

\ArticleDates{Received February 12, 2016, in f\/inal form May 06, 2016; Published online May 12, 2016}

\Abstract{For the two-parameter $p,q$-deformed Heisenberg algebra introduced
 recently and in which, instead of usual commutator of $X$ and $P$ in the l.h.s.\
 of basic relation $[X,P] = {\rm i}\hbar$, one uses the $p,q$-commutator, we
 established interesting properties. Most important is the realizability of
 the $p,q$-deformed Heisenberg algebra by means of the appropriate deformed oscillator algebra.
 Another uncovered property is special extension of the usual mutual Hermitian
 conjugation of the creation and annihilation operators, namely the so-called
 $\eta(N)$-pseudo-Hermitian conjugation rule, along with the related
 $\eta(N)$-pseudo-Hermiticity pro\-perty of the position or momentum operators.
 In this work, we present some new solutions of the realization problem yielding
 new (nonstandard) deformed oscillators, and show their inequivalence to the
 earlier known solution and the respective deformed oscillator algebra,
 in particular what concerns ground state energy.}


\Classification{81R50; 81S05; 81Vxx; 17B37}

\vspace{-2mm}

\section{Introduction}

The obtaining and analysis of modif\/ied/generalized versions of the
 famous Heisenberg uncertainty relation or principle corresponds
 to diverse generalizations of the standard Heisenberg algebra (HA)
 with basic relation $[X, P] = {\rm i}\hbar$ for the position
 and momentum operators. This direction of research is under development for
 more than three decades (with some early works
 quoted as \cite{Brod,Ch-K,Jan,Kempf,Plyush,Saa,Wess}),
 and still remains to be a hot topic,
 see, e.g., \cite{Bagchi,Leiva,Dorsch,GK,GK-2,Maslow}
 for more recent papers.
 Physically, deformations of the Heisenberg algebra and the
 related generalized uncertainty relation f\/ind their motivation
 in quantum gravity, in string theory, noncommutative geometry etc.
 An overview of dif\/ferent approaches can be found, e.g., in \cite{Garay, Hossen}.
 What concerns the particular existing variants of deformed Heisenberg algebra~(DHA),
 we have to note that predominantly there were DHAs with deformed r.h.s.\ of the basic
 relation for $X$ and $P$ that involved a function of momentum,
 or some function of the Hamiltonian. However, within more exotic approach followed
 in \cite{Ch-K,Wess}, deformation of HA emerged in the l.h.s.\ of basic
 relation, by using ``$q$-commutator'' instead of the usual commutator.

 Recently, a hybrid so-called ``two-sided'' deformation
 of HA has been introduced~\cite{GK}.
 In that DHA, f\/irst, the commutator in the l.h.s.\ of its
 def\/ining relation is replaced with $q$-commutator like in~\cite{Ch-K, Wess}
 or with $q,p$-commutator as in~\cite{GK}. In addition,
 the r.h.s.\ of basic relation is modif\/ied by an extra term
 involving Hamiltonian times a pre-factor denoted as $\mu$.
 For this two-sided or $q,p,\mu$-deformed DHA it was shown that,
 similarly to the ``left-handed'' $q$- and $q,p$-deformed ones, this
 more general deformed algebra of the operators $X$ and~$P$
 can as well be realized by (i.e., mapped onto)
 def\/inite nonstandard deformed oscillator algebra (DOA).

 Unusual thing concerning just the $q,p,\mu$-DHA is that
 the deformation factor (parameter) $\mu$ in front of the Hamiltonian $H$
 in the r.h.s.\ of basic relation becomes {\em inevitably} depending not only
 on the deformation parameter(s) $q$ or $q,p$ of the $q$- or $q,p$-commutator
 in the l.h.s., but also on the particle number operator $N$, see~\cite{GK}.

 The explicit connection of the DHA with the respective DOA discloses
 an unexpected pro\-per\-ty~\cite{GK-2} of the operators involved.
 We mean the option that the creation and annihilation
 opera\-tors~$a^+$ and~$a^-$ may be not the (mutual) Hermitian
 conjugates, but the so-called $\eta$-pseudo-Hermitian conjugates
 of one another, with $\eta \equiv \eta(N)$~-- certain operator function
 of the number operator~$N$.
 Accordingly, the position operator $X$ or the momentum operator~$P$,
 or both, were shown to possess not the customary Hermiticity,
 but the property of being $\eta(N)$-pseudo-Hermitian~\cite{GK-2}.

 The latter fact was uncovered in the framework of
 the particular DOA presented explicitly through its structure
 function of deformation (DSF) $\Phi(N)$.
 That was found in closed form in~\cite{GK}.
 However, further analysis shows that this solution given by the DSF~$\Phi(N)$
 is not unique, and other possibilities may exist.
 The goal of this paper is to present other admissible solutions
 which as well provide realization of DHA by some (dif\/ferent) DOA.
 After presenting the new solutions, we demonstrate that these are
 inequivalent to the f\/irst one, found in~\cite{GK} and given by~$\Phi(N)$.
 As a major feature, they show dif\/fering values of the ground state energy.
 The latter can be either lower or higher than the familiar ground state energy
 $E_0=\frac12\hbar\omega$ of the usual quantum harmonic oscillator\footnote{Note
 that this same ground state energy $E_0=\frac12\hbar\omega$ is
 shared by such well-known deformed oscillators as Arik--Coon (AC)
 $q$-oscillator~\cite{Arik}, Biedenharn--Macfarlane (BM)
 $q$-oscillator~\cite{Bied,Mcf}, Tamm--Dancof\/f (TD) type
 $q$-oscillator~\cite{T-D}, and the $p,q$-deformed
 oscillator~\cite{Arik-2, Chakra}.}.

\section[$q$-deformed Heisenberg algebra]{$\boldsymbol{q}$-deformed Heisenberg algebra}\label{section2}

An alternative approach to deform the Heisenberg algebra was
considered in the literature.
 Namely, in~\cite{Ch-K, Wess}
 dif\/ferent approach of deforming HA was studied
 such that a deformation is introduced in the commutator
 in the l.h.s.\ of basic relation (this is the $q$-deformed HA)
\begin{gather} \label{Kl}
X P-q PX = {\rm i}\hbar .
 \end{gather}
 For convenience, in all our treatment we put $\hbar = 1$.
 Below, in the f\/irst subsection we mainly follow~\cite{Ch-K}.
 We require that the equality~(\ref{Kl}) is connected
 with special deformed oscillator algebra whose generating
 elements~$a^+$,~$a^-$ and~$N$ (the creation/annihilation operators,
 which are not necessarily strict conjugates of each other, and the excitation
 number operator) satisfy
\begin{gather} \label{N-a}
 [N,a^+]=a^+ , \qquad [N,a^-]=-a^- ,
\\ \label{H-G}
H(N)a^- a^+ - G(N)a^+ a^- = 1 .
\end{gather}
It is meant that the operator functions $H(N)$ and $G(N)$ admit
formal power series expansion.

\subsection{From DHA to DOA}\label{section2.1}

 Like in \cite{Ch-K} we express the position and momentum operators in terms
 of $a^-$, $a^+$ as
\begin{gather} \label{f-g}
X \equiv f(N) a^- + g(N) a^+ , \qquad P \equiv {\rm i}\big(k(N) a^+ - h(N) a^-\big)
\end{gather}
where $f(N)$, $g(N)$, $h(N)$, $k(N)$ are some functions of the
operator~$N$.

On the base of (\ref{N-a}), for any function
 ${\cal F}(N)$ possessing formal power series expansion we have
\begin{gather} \label{F(N)}
{\cal F}(N) a^{\pm} = a^{\pm} {\cal F}(N\pm 1), \qquad [{\cal F}(N),a^\pm a^\mp] =0.
\end{gather}
Proceeding like in~\cite{Ch-K} and using (\ref{H-G})--(\ref{F(N)}), we deduce
\begin{gather} \label{h/h}
\frac{ h(N+1)}{ h(N)} = q \frac{ f(N+1)}{ f(N)} , \qquad
 \frac{ k(N-1)}{ k(N)} = {q} \frac{ g(N-1)}{ g(N)}
\end{gather}
as well as the expressions
\begin{gather} \label{H(N)}
H(N) = f(N) k(N+1) + q h(N) g(N+1) ,\\
\label{G(N)} G(N) = g(N) h(N-1) + q k(N) f(N-1) .
\end{gather}
The DSF $\Phi(N)$,
 see, e.g.,~\cite{Melj},
 determines both the bilinears (cf.~(\ref{F(N)}))
\begin{gather*}
a^+ a^- = \Phi(N) , \qquad a^- a^+ = \Phi(N+1) ,
\end{gather*}
 and the commutation relation
 \begin{gather} \label{a-a+}
[a^-, a^+]=\Phi(N+1)-\Phi(N) .
\end{gather}
It also gives the action formulas in the $\Phi(n)$-deformed analog of Fock space
\begin{gather*} N |n\rangle =n |n\rangle, \qquad |n\rangle =\frac{(a^+)^n}{\sqrt{\Phi(n)!}} |0\rangle , \qquad a^-|0\rangle = 0,
\end{gather*}
where $\Phi(n)! \equiv\Phi(n)\Phi(n-1)\cdots \Phi(2) \Phi(1)$ and,
in addition, $\Phi(0)!=1$.
 In that space we have
\begin{gather} \label{a+n}
a^+ |n\rangle = \sqrt{\Phi(n+1)} |n+1\rangle , \qquad a^-|n\rangle = \sqrt{\Phi(n)} |n-1\rangle .
\end{gather}
The DSF is obtainable from the above functions
 $H(n)$ and $G(n)$ using the recipe~\cite{Melj}
\begin{gather} \label{Phi-GH}
{\Phi(n)} = \frac{G(n-1)!}{H(n-1)!}\left(\frac{1}{H(0)} + \sum^{n-1}_{j=1} \frac{H(j-1)!}{G(j)!} \right) .
\end{gather}
Here the factorials are def\/ined similarly to $\Phi(n)!$.

\subsection{Solutions of the relations (\ref{h/h})}\label{section2.2}

 We need the solutions of (\ref{h/h}) which then, using (\ref{H(N)}) and (\ref{G(N)}),
 yield the corresponding operator functions $G(N)$ and $H(N)$.
 So let us list some of them.

\emph{Solution}~{\bf A}.
 This is
\begin{gather} \label{f,k(N)}
f(N)= k(N)=\frac{1}{\sqrt 2} {q}^{N},
 \qquad
 h(N)= g(N)= \frac{1}{\sqrt 2} {q}^{2N} ,
\end{gather}
 from which
 \begin{gather} \label{H(N)a}
 H(N) = \frac12 {q}^{2N+1}\big(1
 + q^{2N+2} \big) , \qquad
G(N) = \frac12 {q}^{2N}\big(1 + {q}^{2N-2} \big)
 = q^3 H(N-2) .
\end{gather}
From (\ref{Phi-GH}) and (\ref{H(N)a}) we obtain the DSF
\begin{gather} \label{Phi(n)}
{\Phi_q^{(1)}(n)} = \frac{2 q^{-n}}{(1+q^{2n-2})
(1+q^{2n})}\left(1+\frac{q^{n}-q^{-n+1}}{q-1}\right)
 = \frac{2 q^{-n} [n]_q (1+q^{-n+1}) }{(1+q^{2n-2}) (1+q^{2n})},
 \end{gather}
 where $[n]_q\equiv (1-q^n)/(1-q)$.
 The obtained DSF~(\ref{Phi(n)}) implies that now we have, besides the relation~(\ref{H-G}), also the alternative form~(\ref{a-a+}) of the commutation relation.

In addition, putting (\ref{f,k(N)}) into (\ref{f-g}),
for the operators $X$ and $P$ we obtain
 \begin{gather} \label{XP-1}
X=\frac{1}{\sqrt 2} \bigl(q^{2N} a^+ + q^N a^- \bigr) ,
\qquad
 P=\frac{\rm i}{\sqrt 2} \bigl(q^N a^+-q^{2N} a^-\bigr).
 \end{gather}
Note that this DSF~\eqref{Phi(n)} coincides with the deformation structure function
formerly found in~\cite{GK} (see also~\cite{Ch-K}). However, new solutions are possible,
see the cases ${\bf B}$--${\bf D}$.

\emph{Solution} {\bf B}. This has the form
\begin{gather} \label{f,k(N)-2}
f(N)= k(N)=\frac{1}{\sqrt 2} {q}^{-2N} ,
 \qquad
 h(N)= g(N)= \frac{1}{\sqrt 2} {q}^{-N} ,
\end{gather}
which then yields
\begin{gather} \label{H(N)b}
 H(N) = \frac12 {q}^{-2N}\big(1 + q^{-2N-2} \big) ,
 \qquad
G(N) = \frac12 {q}^{-2N+1}\big(1 + {q}^{-2N+2} \big)
 = q^{-3} H(N-2) .\!\!\!
\end{gather}
Then from (\ref{Phi-GH}) and (\ref{H(N)b}) we obtain the DSF (which
casts (\ref{H-G}) into the form (\ref{a-a+}))
\begin{gather} \label{Phi(n)-2}
 {\Phi_q^{(2)}(n)} = \frac{2 q^{5n-3} [n]_q (1+q^{-n+1}) }
 {(1+q^{2n-2})(1+q^{2n})}
 = q^{3(2n-1)} {\Phi_q^{(1)}(n)} .
\end{gather}
 Moreover, putting (\ref{f,k(N)-2}) into (\ref{f-g}),
for the operators $X$ and $P$ we obtain
\begin{gather} \label{XP-2}
X=\frac{1}{\sqrt 2} \big(q^{-N} a^+ + q^{-2N} a^- \big) ,
 \qquad
 P=\frac{\rm i}{\sqrt 2} \big(q^{-2N} a^+-q^{-N} a^-\big).
 \end{gather}

\emph{Solution} {\bf C}. In this case we f\/ind
\begin{gather} \label{f,k(N)-3}
f(N)= \frac{1}{\sqrt 2} {q}^{-2N}, \qquad
k(N)=\frac{1}{\sqrt 2} {q}^{N} , \qquad
 g(N)= \frac{1}{\sqrt 2} {q}^{2N} , \qquad
 h(N)= \frac{1}{\sqrt 2} {q}^{-N},
\end{gather}
from which we infer
\begin{gather} \label{H(N)c}
H(N) = \frac12 {q}^{-N+1}\big(1
 + q^{2N+2} \big) ,
 \qquad
G(N) = \frac12 {q}^{N+1}\big(1 + {q}^{-2N+2} \big) = H(N-2) .
\end{gather}
From (\ref{Phi-GH}) and (\ref{H(N)c}) we obtain the DSF (which casts~(\ref{H-G}) into the form~(\ref{a-a+}))
\begin{gather} \label{Phi(n)-3}
{\Phi_q^{(3)}(n)} = \frac{2 q^{2n-3} [n]_q (1+q^{-n+1}) }
 {(1+q^{2n-2})(1+q^{2n})} =q^{3(n-1)} {\Phi_q^{(1)}(n)} .
\end{gather}
%%%
In addition, putting (\ref{f,k(N)-3}) into (\ref{f-g}),
 for the operators $X$ and $P$ we f\/ind
\begin{gather} \label{XP-3}
X=\frac{1}{\sqrt 2} \big(q^{2N} a^+ + q^{-2N} a^- \big) ,\qquad
 P=\frac{\rm i}{\sqrt 2} \big(q^N a^+-q^{-N} a^-\big).
 \end{gather}

\emph{Solution} {\bf D}. That reads
\begin{gather} \label{f,k(N)-4}
f(N)= \frac{1}{\sqrt 2} {q}^{N}, \qquad
k(N)=\frac{1}{\sqrt 2} {q}^{-2N} , \qquad
 g(N)= \frac{1}{\sqrt 2} {q}^{-N} , \qquad
 h(N)= \frac{1}{\sqrt 2} {q}^{2N},
\end{gather}
so that
 \begin{gather} \label{H(N)d}
H(N) = \frac12 {q}^{N}\big(1 + q^{-2N-2} \big) , \qquad
G(N) = \frac12 {q}^{-N}\big(1 + {q}^{2N-2} \big)= H(N-2) .
\end{gather}
From (\ref{Phi-GH}) and (\ref{H(N)d}) we obtain the DSF (which casts
(\ref{H-G}) into the form (\ref{a-a+}))
\begin{gather} \label{Phi(n)-4}
{\Phi_q^{(4)}(n)} = \frac{2 q^{2n} [n]_q (1+q^{-n+1}) }
 {(1+q^{2n-2})(1+q^{2n})}
= q^{3n} {\Phi_q^{(1)}(n)} = q^3 {\Phi_q^{(3)}(n)} .
\end{gather}
At last, putting (\ref{f,k(N)-4}) into (\ref{f-g}),
 for the operators $X$ and $P$ we obtain
 \begin{gather} \label{XP-4}
X=\frac{1}{\sqrt 2} \big(q^{-N} a^+ + q^N a^- \big) ,
 \qquad
 P=\frac{\rm i}{\sqrt 2} \big(q^{-2N} a^+-q^{2N} a^-\big).
 \end{gather}

\begin{remark}\label{remark1}
 Each of the solutions {\bf A}--{\bf D} in the limit $q\to 1$ yields
 $f=g=h=k=1/{\sqrt 2}$, $G=H=1$.
 Then we recover the structure function $\Phi(n)=n$ of the usual oscillator,
 along with known relations $X=(a^++a^-)/{\sqrt 2}$,
 $P = {\rm i}(a^+ -a^-)/{\sqrt 2}$.
 \end{remark}

 \begin{remark}[concerning pseudo-Hermiticity]\label{remark2}
 From any of the relations (\ref{XP-1}), (\ref{XP-2}),
 (\ref{XP-3}), (\ref{XP-4}) in can be
 deduced that the position operator and the momentum operator cannot be
the both Hermitian simultaneously. Say, using the relation identical
to (\ref{XP-1}) it has been shown in~\cite{GK-2} that (i)~if $P$ is
chosen
 to be Hermitian, then~$X$ turns out to be $\eta_X(N)$-pseudo-Hermitian;
 (ii)~if $X$ is f\/ixed to be Hermitian, then $P$ is $\eta_P(N)$-pseudo-Hermitian;
 (iii)~in general, the both of~$X$,~$P$ are non-Hermitian and, moreover, $X$ is
 $\eta_X(N)$-pseudo-Hermitian and $P$ is $\eta_P(N)$-pseudo-Hermitian.
 For more details see~\cite{GK-2}.
 \end{remark}

\begin{figure}[t]\centering
\includegraphics[width = 0.6 \linewidth]{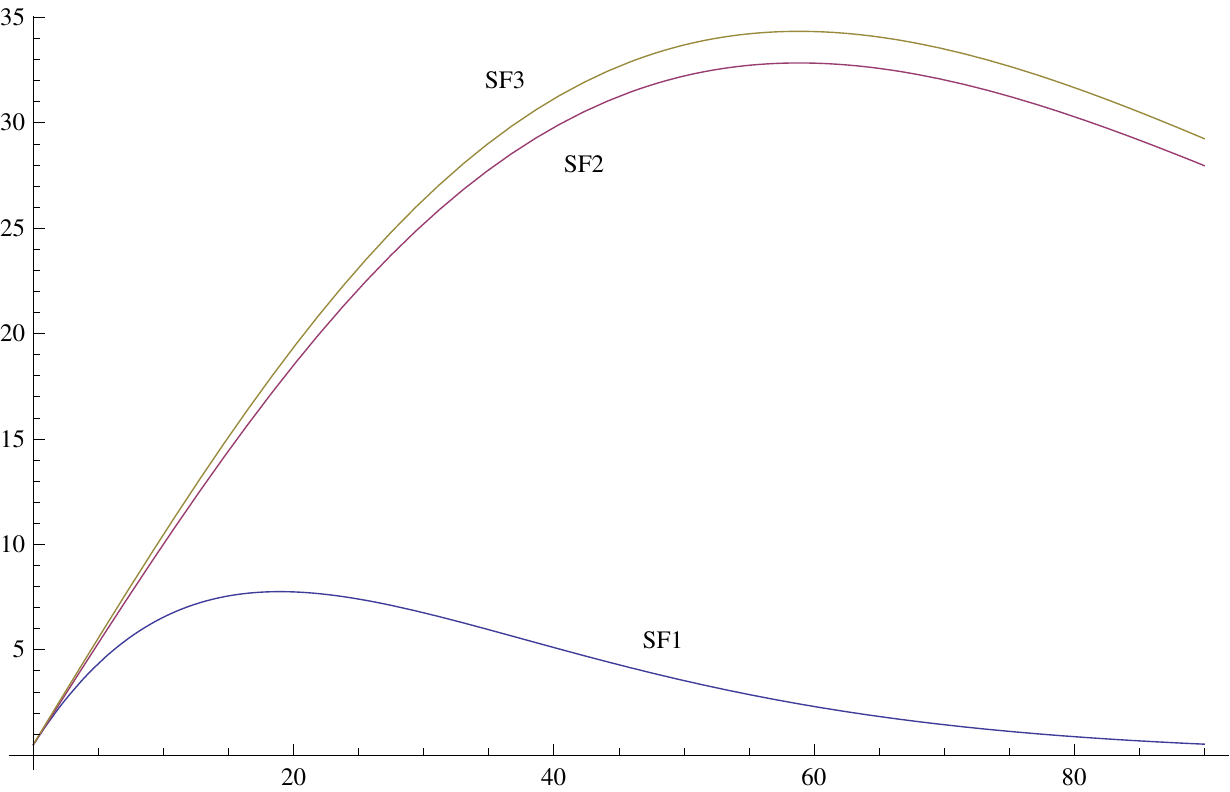}
\caption{Deformation structure functions versus $n$ at f\/ixed value
$q=1.015$: here SF1, SF2 and SF3 denote respectively
$\Phi_q^{(1)}(n)$, $\Phi_q^{(2)}(n)$ and $\Phi_q^{(3)}(n)$ from
(\ref{XP-1}), (\ref{XP-2}) and (\ref{XP-3}).} \label{fig1}
\end{figure}

 Now let us examine the behavior of all the structure
functions as ref\/lected in the energy eigenspectrum.
 With the Hamiltonian and the energy eigenspectrum
 given as
\begin{gather*} %begin{equation}
 {\cal H}=\frac12 (a a^+ + a^+ a)=
 \frac12 \bigl(\Phi_q(N+1)+ \Phi_q(N)\bigr),
\qquad
 E_q(n)= \frac12
\bigl(\Phi_q(n+1)+ \Phi_q(n)\bigr) ,
 \end{gather*}
 we obtain the energies $E(n)$ for the respective DSFs.
 In Fig.~\ref{fig1}, the structure functions $\Phi_q^{(1)}(n)$, $\Phi_q^{(2)}(n)$ and
$\Phi_q^{(3)}(n)$ from~(\ref{XP-1}), (\ref{XP-2}) and
 (\ref{XP-3}) are shown for the
 particular value $q=1.015$ of the deformation parameter.
 Note that the 4th DSF $\Phi_q^{(4)}(n)$,
 due to the simple relation with $\Phi_q^{(3)}(n)$, see~(\ref{Phi(n)-4}), and
 the chosen $q=1.015$, looks almost the same as $\Phi_q^{(3)}(n)$ and so
 is not shown in Fig.~\ref{fig1}.

\begin{remark}[ground state energies]\label{remark3}
 It is of interest to examine the ground state (at $n=0$) energy
 eigenvalues for each of the cases {\bf A}--{\bf D}.
 As seen from (\ref{Phi(n)}), (\ref{Phi(n)-2}), (\ref{Phi(n)-3}), (\ref{Phi(n)-4}),
 each of the DSFs $\Phi_q^{(1)}(n)$, $\ldots$ , $\Phi_q^{(4)}(n)$
 give zero at $n=0$, due to the common factor $[n]_q$ such that $[0]_q=0$.
 Therefore the ground state energy is given by (one half of) the
 corresponding DSF value at $n=1$.
 This way we obtain the following data for $E_q^{(j)}(0)$
 \begin{alignat*}{3}
 & E_q^{(1)}(0)=\frac12 \Phi_q^{(1)}(1)=\frac{q^{-1}}{1+q^2} ,
 \qquad &&
 E_q^{(2)}(0)=\frac12 \Phi_q^{(2)}(1)=\frac{q^2}{1+q^2} ,&
\\
& E_q^{(3)}(0)=\frac12 \Phi_q^{(3)}(1)=\frac{q^{-1}}{1+q^2} ,
 \qquad &&
 E_q^{(4)}(0)=\frac12
 \Phi_q^{(4)}(1)=\frac{q^2}{1+q^2} . &
\end{alignat*}
 Assuming $q>1$ (like in Fig.~\ref{fig1}) we f\/ind that, in comparison
 with the zero-level energy $E_{q}(0)|_{q=1}= \frac12$ of the usual quantum
 oscillator, here we have both the increased and the lowered ground state
 energies of deformed oscillators
 \begin{gather*} E_q^{(1)}(0) = E_q^{(3)}(0) < \frac12,
 \qquad
 {\rm whereas}
 \quad
 E_q^{(2)}(0) = E_q^{(4)}(0) > \frac12 .
 \end{gather*}
 On the other hand, if $q < 1$ we f\/ind
 \begin{gather*} E_q^{(1)}(0) = E_q^{(3)}(0) > \frac12,
 \qquad
 {\rm while}
 \quad
 E_q^{(2)}(0) = E_q^{(4)}(0) < \frac12 .
 \end{gather*}
 \end{remark}

\begin{remark}[concerning accidental degeneracy]\label{remark4}
 As seen from the behavior of the structure functions
 pictured in Fig.~\ref{fig1}, the considered deformed oscillators may possess,
 at certain corresponding values of the deformation parameter~$q$,
 diverse cases of accidental pairwise energy level
 degeneracy (note that such kind of degeneracy in one dimension
 is peculiar for certain class of {\it deformed} oscillators
 that was formerly studied in~\cite{GR-1,GR-2,GR-3},
 along with the modif\/ied versions in~\cite{C-G-K-R,G-K-R, GR-UJP}).
 For instance, for the deformed oscillator given
 by the DSF $\Phi_{q}^{(1)}(n)$ from (\ref{Phi(n)}) there exists
 certain value $q=q(n)|_{n=2}$ such that the degeneracy $E_q(0)=E_q(2)$ is realized.
 In the case of general $n$ such value is to be found by solving the equation
\begin{gather*}
E_q^{(1)}(n) - E_q^{(1)}(0)
 = \frac12 \big[ \Phi_q^{(1)}(n+1) + \Phi_q^{(1)}(n) \big] -
\frac{q^{-1}}{1+q^2}=0
 \end{gather*}
 or, in a more expanded form, the equation
 \begin{gather*}
 q^{4 n}\big(q^2+q^{-2}\big) + q^{3 n}(q-1)\big(q^2+q^{-3}\big)
 -q^{2 n}\big(q^3+q^{-3}-2\big)\\
 \qquad{}+q^n \big(q-q^{-1}\big)-q-q^{-1}-\frac{1-q^{-1}}{1+q^2}q^{2
 n}\big(1+q^{2n-2}\big)\big(1+q^{2n}\big)\big(1+q^{2n+2}\big)=0 .
 \end{gather*}
 For instance, if one f\/ixes $n=10$, the solution for $E_q^{(1)}(10)-E_q^{(1)}(0)=0$
 is $q_{10} = 1.0913$. Likewise, if $n=90$ the degeneracy $E_q^{(1)}(90)= E_q^{(1)}(0)$
 occurs at $q_{90} = 1.015148$. Note that~$q_{90}<q_{10}$, i.e., the larger is $n$ the
 lesser is respective value~$q_n$.

 In a similar fashion, for each of the DSFs $\Phi^{(j)}(n)$, $j=1,2,3,4$,
 one can consider these and many other cases of degeneracies
 (and f\/ind the relevant values of $q$), e.g., such as $E_q(1)=E_q(7)$,
 $E_q(2)=E_q(5)$, the neighboring $E_q(3)=E_q(4)$ and so on.
 \end{remark}

 \subsection{General approach to the relations (\ref{h/h})}\label{section2.3}

 Let us take (\ref{h/h}) in the equivalent form
 \begin{gather*}
 \frac{h(n+1)}{f(n+1)} = q \frac{h(n)}{f(n)} .
 \end{gather*}
The latter, by denoting the ratio as $\frac{h(n)}{f(n)}\equiv d(n)$,
implies $d(n+1)=q d(n)$ from which we infer $d(n)=q^n d(0)$.
 That yields the relation
 %With the choice $d(0)=1$
 \begin{gather*}
h(n)= q^n f(n) d(0) ,
\end{gather*}
and similarly (putting $ c(0)\equiv {g(0)}/{k(0)}$) we f\/ind the
relation
\begin{gather*}
g(n) = q^n k(n) c(0) .
\end{gather*}
With account of the latter two relations, from (\ref{H(N)}) and
(\ref{G(N)}) we infer the desired functions $G(N)$ and $H(N)$:
 \begin{gather} \label{G-gen}
 G(N) = q f(N-1) k(N) \bigl(1 + c(0) d(0) q^{2N-2}\bigr) ,
\\ \label{H-gen}
 H(N) = f(N) k(N+1) \bigl(1 + c(0) d(0) q^{2N+2} \bigr) ,
 \end{gather}
along with
\begin{gather*} H(0) = f(0) k(1) \bigl(1+c(0)d(0)q^2\bigr)
\end{gather*}
(of course, there is natural simplifying choice $c(0)=d(0)=1$).
 The expressions~(\ref{G-gen}) and~(\ref{H-gen}) can be used for
 obtaining the corresponding DSF.

 By introducing ${\cal R}(N)\equiv f(N-1) k(N)$
 we can present~(\ref{G-gen}) and~(\ref{H-gen}) in the form
 \begin{gather*} G(N) = q {\cal R}(N) \bigl(1 + c(0) d(0) q^{2N-2}\bigr) ,
 \qquad
 H(N) = {\cal R}(N+1) \bigl(1 + c(0) d(0) q^{2N+2}\bigr) ,
 \end{gather*}
with the condition $G(N)|_{q\to 1}=H(N)|_{q\to 1}=1$, see~(\ref{H-G}) and Remark~\ref{remark1}. But that means ${\cal R}(N)={\cal
R}(N+1)$ if $q=1$, or ${\cal R}(N)|_{q\to 1} = {\rm const}$.

\section[A $q,p$-deformed HA and $q,p$-oscillators]{A $\boldsymbol{q,p}$-deformed HA and $\boldsymbol{q,p}$-oscillators}\label{section3}

An extended two-parameter deformation of HA, see~\cite{GK}, obeys
the basic relation
\begin{gather} \label{pqHA}
 p X P-q PX = {\rm i}\hbar , \qquad p\neq q ,
 \qquad p\neq 1 , \qquad q\neq 1.
\end{gather}
 Denote $Q\equiv \frac{q}{p}$.
 In analogy to (\ref{h/h}), (\ref{H(N)}) and (\ref{G(N)})
 we obtain the formulas
\begin{gather} \label{f'/f'}
\frac{\tilde{f}(N+1)}{ \tilde{f}(N)} =
 Q^{-1} \frac{\tilde{h}(N+1)}{ \tilde{h}(N)} ,
 \qquad
\frac{ \tilde{g}(N-1)}{ \tilde{g}(N)} = Q^{-1} \frac{
\tilde{k}(N-1)}{ \tilde{k}(N)} ,
\end{gather}
 and the relations
\begin{gather*}
\tilde{H}(N) = p \tilde{f}(N) \tilde{k}(N+1) + q \tilde{h}(N)
\tilde{g}(N+1) ,\\
\tilde{G}(N) = p \tilde{g}(N) \tilde{h}(N-1) + q \tilde{k}(N)
\tilde{f}(N-1) .
\end{gather*}
Now we seek solutions of~(\ref{f'/f'}).
 For each of them, we also give the corresponding $\tilde{H}(N)$, $\tilde{G}(N)$,
 the operators~$X$,~$P$, and the DSF.

\emph{Solution} ${\bf\tilde{A}}$. The f\/irst solution is
\begin{gather*} % \label{f'(N)-1}
\tilde{f}(N) = \tilde{k}(N) = \frac{1}{\sqrt 2} Q^{N},
\qquad \tilde{h}(N) = \tilde{g}(N)= \frac{1}{\sqrt 2}Q^{2N}
\end{gather*}
 that leads to the result
\begin{gather} \label{HG'-1}
\tilde{H}(N) = \frac12 p Q^{2N+1}\big( 1 + Q^{2N+2} \big) ,
 \qquad \tilde{G}(N) = \frac12 p Q^{2N}\big( 1 + Q^{2N-2} \big)= Q^3 \tilde{H}(N-2) ,
\end{gather}
 along with
\begin{gather*}
X=\frac{1}{\sqrt 2} \big[Q^{2N} a^+ + Q^N a^- \big] , \qquad P=\frac{\rm i}{\sqrt 2} \big[Q^N a^+ - Q^{2N} a^-\big] .
 \end{gather*}
 The DSF $\Phi_{q,p}(N)$ (it determines the relation of $a^\pm a^\mp$
 with~$N$ like in~(\ref{a-a+}) as well as the action formulas
 for~$a^+$,~$a^-$, see~(\ref{a+n})) is inferred from equation~(\ref{Phi-GH})
 using the functions~$\tilde{H}(n)$ and~$\tilde{G}(n)$ in~(\ref{HG'-1}).
 The result is
\begin{gather}
{\Phi_{q,p}^{(1)}(n)} = \frac{2 p^{-1} Q^{-n}}{(1+Q^{2n-2})
(1+Q^{2n})} \left(1+\frac{Q^{n}- Q^{-n+1}}{Q-1}\right) \nonumber\\
\label{Phi'(n)}
 \hphantom{{\Phi_{q,p}^{(1)}(n)}}{} = \frac{2 q^{-n} p^{5n-3}}
 {(q^{2n-2}+p^{2n-2})(q^{2n}+p^{2n})}
 \left(1+\frac{[2n-1]_{q,p}}{(qp)^{n-1}}\right)
 = \frac{2 p^{-1} Q^{-n} [n]_Q (1+Q^{1-n})}
 { (1+Q^{2n-2})(1+Q^{2n})},\!\!\!
\end{gather}
 where
\begin{gather*} [x]_{q,p}\equiv \frac{p^x-q^x}{p-q} , \qquad [x]_{Q}\equiv \frac{1-Q^x}{1-Q}.
\end{gather*}
 This DSF determines the nonstandard two-parameter deformed oscillator
 coinciding with that found in~\cite{GK}.
 It is ($q\leftrightarrow p$)-nonsymmetric and thus obviously
 dif\/fers from the well known $q,p$-oscillator~\cite{Chakra,Arik-2}
 whose structure function $\Phi_{q,p}(n)=[n]_{q,p}$ is
($q\leftrightarrow p$)-symmetric.

\emph{Solution} ${\bf\tilde{B}}$.
 The next solution is
\begin{gather*} %\label{f'(N)-1}
\tilde{f}(N) = \tilde{k}(N) = \frac{1}{\sqrt 2}
Q^{-2N} , \qquad \tilde{h}(N) = \tilde{g}(N)=\frac{1}{\sqrt 2} Q^{-N},
\end{gather*}
 that leads to the result
\begin{gather}
\tilde{H}(N) = \frac12 p Q^{-2N}\bigl( 1 + Q^{-2N-2} \bigr) ,\nonumber\\
 \label{HG'-2} \tilde{G}(N)=\frac12 p Q^{-2N+1}\bigl( 1 + Q^{-2N+2}\bigr)=Q^{-3}\tilde{H}(N-2) ,
\end{gather}
 along with
\begin{gather*}
X=\frac{1}{\sqrt 2} \big[Q^{-N} a^+ + Q^{-2N} a^- \big] ,\qquad P=\frac{\rm i}{\sqrt 2} \big[Q^{-2N} a^+ - Q^{-N} a^-\big] .
 \end{gather*}
 The DSF $\Phi_{q,p}(N)$ (which determines the relation of $a^\pm a^\mp$
 with $N$ like in (\ref{a-a+}) as well as the action formulas
 for $a^+$, $a^-$, see~(\ref{a+n})) is inferred from equation~(\ref{Phi-GH})
 using the functions $\tilde{H}(n)$ and $\tilde{G}(n)$ in~(\ref{HG'-2}).
 The result is
 \begin{gather} \label{Phi'(n)-2}
\Phi_{q,p}^{(2)}(n) = \left(\frac{q}{p}\right)^{3(2n-1)} \Phi_{q,p}^{(1)}(n) .
\end{gather}
 This DSF determines the second, nonstandard ($q \leftrightarrow p$)-nonsymmetric two-parameter deformed oscillator.

\emph{Solution} ${\bf\tilde{C}}$.
 The next solution is
\begin{gather} % \label{f'(N)-1}
\tilde{f}(N) = \frac{1}{\sqrt 2} Q^{-2N} , \qquad
\tilde{k}(N)=\frac{1}{\sqrt 2} Q^{N} ,\qquad
 \tilde{g}(N)= \frac{1}{\sqrt 2} Q^{2N} , \qquad
 \tilde{h}(N)= \frac{1}{\sqrt 2} Q^{-N},
\end{gather}
 that leads to the result
 \begin{gather} \label{HG'-3}
\tilde{H}(N) = \frac12 p Q^{1-N}\bigl( 1 + Q^{2N+2} \bigr) ,
 \qquad
\tilde{G}(N) = \frac12 p Q^{N+1}\bigl( 1 + Q^{-2N+2} \bigr)=\tilde{H}(N-2) ,
\end{gather}
 along with
\begin{gather*}
X=\frac{1}{\sqrt 2} \bigl[Q^{2N} a^+ + Q^{-2N} a^- \bigr] , \qquad P=\frac{\rm i}{\sqrt 2} \bigl[ Q^N a^+ - Q^{-N} a^-\bigr] .
 \end{gather*}
 The DSF $\Phi_{p,q}(N)$ (it provides the relation of $a^\pm a^\mp$
 with $N$ like in~(\ref{a-a+}) and the action formulas
 for~$a^+$,~$a^-$, see~(\ref{a+n})) is inferred from equation~(\ref{Phi-GH})
 using the functions $\tilde{H}(n)$ and $\tilde{G}(n)$ from~(\ref{HG'-3}). The result is{\samepage
\begin{gather} \label{Phi'(n)-3}
\Phi_{q,p}^{(3)}(n) = \left(\frac{q}{p}\right)^{3(n-1)} \Phi_{q,p}^{(1)}(n) .
\end{gather}
This DSF determines yet another nonstandard, ($q \leftrightarrow p$)-nonsymmetric deformed oscillator.}

\emph{Solution} ${\bf\tilde{D}}$. This solution is of the form
\begin{gather*} %\label{f'(N)-1}
\tilde{f}(N) = \frac{1}{\sqrt 2} Q^{N} , \qquad
\tilde{k}(N)=\frac{1}{\sqrt 2} Q^{-2N} ,\qquad
 \tilde{g}(N)= \frac{1}{\sqrt 2} Q^{-N} , \qquad \tilde{h}(N)= \frac{1}{\sqrt 2} Q^{2N},
\end{gather*}
 that leads to the result
\begin{gather} \label{HG'-4}
\tilde{H}(N) = \frac12 p Q^{N}\bigl( 1 + Q^{-2N-2} \bigr) , \qquad
\tilde{G}(N) = \frac12 p Q^{-N}\bigl( 1 + Q^{2N-2} \bigr) =\tilde{H}(N-2) ,\\
X=\frac{1}{\sqrt 2} \bigl[Q^{-N} a^+ + Q^N a^- \bigr] , \qquad
 P=\frac{\rm i}{\sqrt 2} \bigl[ Q^{-2N} a^+ - Q^{2N} a^-\bigr] .\nonumber
 \end{gather}
 The DSF $\Phi_{q,p}(N)$ (it determines the relation of~$a^\pm a^\mp$
 with $N$ like in (\ref{a-a+}), and the action formulas
 for~$a^+$, $a^-$, see~(\ref{a+n})) is inferred using equation~(\ref{Phi-GH})
 with the functions~$\tilde{H}(n)$ and~$\tilde{G}(n)$ from~(\ref{HG'-4}).
 The result is
 \begin{gather} \label{Phi'(n)-4}
\Phi_{q,p}^{(4)}(n) = \left(\frac{q}{p}\right)^{3n} \Phi_{q,p}^{(1)}(n) .
\end{gather}
 The latter DSF determines the forth, nonstandard, two-parameter deformed oscillator obviously nonsymmetric under $q\leftrightarrow p$.

Note that at $p\to 1$ the results obtained here
 for the $p,q$-deformed HA~(\ref{pqHA}) reduce to those of the
 preceding section (say, (\ref{Phi'(n)}) reduces to (\ref{Phi(n)}), etc.),
 whereas for the case $p=q\neq 1$ we come to the
 structure function $\phi(n)=\frac{n}{q}$,
 the familiar operators $X=\frac{1}{\sqrt{2}}(a^+ + a^-)$ and
 $P=\frac{\rm i}{\sqrt{2}}(a^+ - a^-)$ along with $[a^-,a^+]=1/q$ and $\tilde{H}(N)=H=q$, $\tilde{G}(N)=G=q$.
 Obviously, we deal again with the usual harmonic oscillator,
 but the spacing in its (linear) energy spectrum is $\frac1q$-scaled.

\begin{remark}\label{remark5}
 Using the two-parameter family of
 ($p \leftrightarrow q$)-symmetric $p,q$-oscillators from~\cite{Arik-2, Chakra}
 one can infer, see~\cite{GR-3}, a whole ``plethora''
 of one-parameter $q$-deformed oscillators.
 This variety includes such well-known or ``standard'' ones
 as Biedenharn--Macfarlane \cite{Bied,Mcf} (if $p=q^{-1}$),
 Arik--Coon \cite{Arik} (if $p=1$), and Tamm--Dancof\/f~\cite{T-D} (if $p=q$) $q$-oscillators.
 Now, quite analogously, by imposing diverse functional dependences $p=\xi(q)$
 it is possible to deduce from each of the new ($p \leftrightarrow q$)-{\it
 nonsymmetric} $\tilde{\bf A}$--$\tilde{\bf D}$ families found in this section,
 see (\ref{Phi'(n)}), (\ref{Phi'(n)-2}), (\ref{Phi'(n)-3}) and~(\ref{Phi'(n)-4}),
 the corresponding alternative ``plethoras'' of {\it non-standard}
 $q$-deformed oscillators, of which only relatively simple examples
 (got by setting $p=1$) are given as the solutions ${\bf A}$--${\bf D}$ above (in Section~\ref{section2.2}).
\end{remark}

\begin{remark}\label{remark6}
 The parameters $p$ and $q$ in the def\/ining relation (\ref{pqHA})
 of the $q,p$-DHA may be either real or complex.
 The issue of which particular (complex) values of $p$, $q$ are
 admissible was discussed in~\cite[Section~4]{GK-2}~-- clearly that
 depends on the adopted rules of (pseudo)Hermiticity of~$X$ and~$P$.
 On the other hand, the deformed oscillators obtained in Section~\ref{section2.2}
 and given by the DSFs~(\ref{Phi(n)}),~(\ref{Phi(n)-2}),~(\ref{Phi(n)-3})
 and~(\ref{Phi(n)-4}) admit only real values of the deformation parameter~$q$.
 However, from each of these DSFs, say ${\Phi_q^{(1)}(n)}$ from~(\ref{Phi(n)}),
 we can construct the related ($q\leftrightarrow q^{-1}$)-symmetric deformed
 oscillator by combining this $q$-DSF with its ``cousin'' ${\Phi_{q^{-1}}^{(1)}(n)}$.
That yields the symmetrized (and factorized) DSF
 \begin{gather*} {\Phi_{\rm symm.}^{(1)}(n)}\equiv \frac12\bigl({\Phi_q^{(1)}(n)} +
{\Phi_{q^{-1}}^{(1)}(n)}\bigr) =
\frac{q^{-n} [n]_q (1+q^{-n+1})}{(1+q^{2n-2})(1+q^{2n})} +
 \frac{q^{n} [n]_{q^{-1}} (1+q^{n-1})}{(1+q^{-2n+2})(1+q^{-2n})}
 \\
 \hphantom{{\Phi_{\rm symm.}^{(1)}(n)}}{}
 =\frac{(q^{3n-1}+q^{-3n+1})(q^{\frac{n-1}{2}}+q^{\frac{-n+1}{2}})}
 {(q^{n}+q^{-n})(q^{n-1}+q^{-n+1})} [[n]]_{q^{1/2}},
 \qquad [[X]]_q\equiv\frac{q^{X}-q^{-X}}{q-q^{-1}} ,
 \end{gather*}
 which admits (real and) the complex form\footnote{For some prof\/its of
 dealing with complex deformation parameter(s) see the last paragraph
 in \cite[Section~7]{C-G-K-R} (with references therein),
 and also the works \cite{GR-2012,Plyush2, Roven}.}
 of $q$, namely $q={\rm e}^{{\rm i}\theta}$, $0 \leq \theta < \pi$.
 The same recipe applies to $\Phi_q^{(j)}(n)$, $j=2,3,4$.
 It is also clear how to proceed in the case of two-parameter
 deformed oscillators of Section~\ref{section3}. Namely, each of those DSFs
 yields the corresponding ($q\leftrightarrow p$)-symmetric
 DSF and its deformed oscillator, by adding
 ${\Phi_{q,p}^{(j)}(n)}$ to its ``cousin'' ${\Phi_{p,q}^{(j)}(n)}$, $j=1,\dots ,4$.
 Then the parameters may be either real or complex such that
 $p=\bar{q}=r{\rm e}^{-{\rm i}\theta}$.

 At last let us note that the DSF ${\Phi_q^{(j)}(n)}$ and its ``cousin'' DSF
 ${\Phi_{q^{-1}}^{(j)}(n)}$ are inferred from dif\/ferent copies of DHA.
 Likewise, ${\Phi_{q,p}^{(j)}(n)}$ and its ``cousin'' ${\Phi_{p,q}^{(j)}(n)}$
 are linked with the (dif\/fering) $q,p$-DHA and $p,q$-DHA respectively.
\end{remark}

\section{Discussion and outlook}\label{section4}

 Our main results are contained in Sections~\ref{section2}, \ref{section3} and give solutions of
 the mapping DHA to DOA problem which allow to present the $q,p$-deformed
 (``left-sided'') HA in terms of respective {\it non-standard}
 deformed oscillators determined through their respective
 structure functions $\Phi^j_{q}(n)$ and $\Phi^j_{q,p}(n)$ where $j=1,2,3,4$.
 Note that the aspects concerning (pseudo-Hermitian) mutual conjugation
 of~$a^-$ and~$a^+$, as well as special non-Hermiticity
 (i.e., $\eta(N)$-pseudo-Hermiticity) of~$X$,~$P$
 can be examined by a detailed analysis, in analogy to what
 was done in~\cite{GK-2}.

 Remembering that in~\cite{GK-2} we used as starting point
 not only $q$- and $p,q$-deformed Heisenberg algebras but also the two-sided
 $p,q,\mu$-deformed DHA, it would be useful to undertake more detailed and
 complete (than in~\cite{GK-2}) study of the problem of f\/inding solutions
 which map the $p,q,\mu$-deformed DHA onto certain
 $p,q,\mu$-deformed oscillators.

 Both the BM-type, AC-type $q$-oscillators and
 the $p,q$-oscillators \cite{Arik,Bied,Chakra,T-D,Mcf}, along with more exotic
 deformed oscillators~\cite{G-K-R,GR-3, Jan}, are used to construct the
 respective one- and two-parameter deformed analogs of Bose gas
 model (see \cite{AdGa,Algin,GM-UJP,GR-2011,GR-2012}
 and references therein) which f\/ind interesting applications
 including phenomenological ones~\cite{SIGMA,GM-NP,Liu}.
 So it is of interest to develop, starting from
 the $q$- or $p,q$-deformed oscillators explored in this paper, the
 corresponding new deformed (certainly {\it non-standard}) models
 with either thermodynamics or statistical distributions
 and correlations in the focus.
 Also, there is an interesting issue of modif\/ied versions
 of the Heisenberg uncertainty relation (with its expected physical
 implications) for the above studied deformations of HA.
 All these topics are worth of detailed future studies.

\subsection*{Acknowledgements}

This work was partially supported by the Special Programme of
Division of Physics and Astronomy of NAS of Ukraine.
The authors also thank the anonymous referees for valuable remarks
 that have led to the improved version of the manuscript.

\pdfbookmark[1]{References}{ref}
\LastPageEnding

\end{document}